\documentclass{elsart}

\usepackage{epsfig}
\usepackage{amsmath}
\usepackage{latexsym}
\usepackage{exscale}

\renewcommand{\vec}[1]{\mbox{\boldmath{$#1$}}}

\begin{document}

\begin{frontmatter}


\title{Silicon Drift Detector Readout Electronics for a Compton Camera \thanksref{caesar}}
\thanks[caesar]{Work partially supported by CAESAR (Center of Advanced European Studies And Research)}

\author[siegenfb7]{T. ~{\c C}onka Nurdan\corauthref{cor}},
\corauth[cor]{Corresponding author. tel: +49 271 7403536; fax: +49 271 7403533}
\ead{conka@alwa02.physik.uni-siegen.de}
\author[siegenfb12]{K.~Nurdan},
\author[siegenfb7]{A.H.~Walenta},
\author[siegenfb7]{H.J.~ Besch},
\author[milano]{C.~Fiorini},
\author[marburg]{B.~Freisleben},
\author[siegenfb7]{N.A.~Pavel}

\address[siegenfb7]{Universit\"at Siegen, FB Physik, Emmy-Noether-Campus, Walter-Flex-Str. 3 57072 Siegen, Germany}
\address[milano]{Politecnico di Milano, Dipartimento di Elettronica e Informazione, Sezione di Elettronica, Via Golgi, 40 20133, Milano, Italy}
\address[siegenfb12]{Universit\"at Siegen, FB Elektrotechnik und Informatik, H\"olderlinstr. 3 57072 Siegen, Germany}
\address[marburg]{Philipps Universit\"at Marburg, FB Mathematik und Informatik, Hans-Meerwein-Str., 35041 Marburg, Germany}



\begin{abstract}
A prototype detector for Compton camera imaging is under development. A monolithic array of 19 channel Silicon drift detector with on-chip electronics is going to be used as a scatter detector for the prototype system. Custom designed analog and digital readout electronics for this detector was first tested by using a single cell Silicon drift detector. This paper describes the readout architecture and presents the results of the measurement. 
\end{abstract}

\end{frontmatter}

\section{Introduction}
Since the introduction of the Compton imaging principle \cite{Todd}, Compton cameras found a number of applications \cite{Schönfelder},\cite{Singh},\cite{Martin} and several prototype detectors have been produced. The prototype system under development which is discussed here \cite{ourIEEEpaper} is going to be used for studying the application in medical imaging. 

The principal idea of the Compton camera is to replace the mechanical collimator of an Anger camera with an electronic collimator. The Compton camera consists of two detector components: the so called scatter detector and the absorption detector. A photon emitted from a source undergoes Compton scattering at the scatter detector where the recoil electron is absorbed and its energy and the location of interaction are determined. The scattered photon leaves the scatter detector and is absorbed in the second detector where the energy and impact position are determined. From this information the source of the incident photon is found to be on the surface of a cone; the so called backprojected cone. 

Silicon was found to be 
the best material for the scatter detector \cite{Ott} considering its high Compton to total interaction ratio at the gamma energy of interest (several hundred keV). The prototype will consist of a Silicon Drift Detector (SDD)\cite{Strüder} as the scatter detector and an Anger camera without lead collimator as the absorption detector. The Silicon detector is a monolithic array of 19 cells with an on-chip JFET for the first amplification in every cell. It has been produced by the Max Planck Institute semiconductor laboratory (MPI/HLL). This paper presents the fast frontend and readout electronics designed for this detector and the test results which were obtained with a single cell detector of the same type.

\section{Setup Overview}

The single cell detector has an area of 5 mm$^2$ and it has a circular shape. The thickness of the wafer is 300 $\mu$m. The mounting of the detector on a ceramic support and the bonding were done at MPI/HLL.
The detector leakage current was around 100 pA. The operational principle of this detector has been explained in several publications elsewhere \cite{MPIHLL}. The most significant feature of this detector is the on-chip integrated JFET which serves to reduce the stray capacitance and therefore provides a better noise performance compared to other Silicon detectors. This detector type has been used for several applications such as a scintillator based gamma-camera \cite{Carlos19SDDpaper}, holography \cite{Holography}, and spectrometry \cite{Spectrometer}. Our prototype Compton camera will exploit the new possible application field for it.

\begin{figure}[htbp]
\centerline
{
 \includegraphics[width=4in,clip]{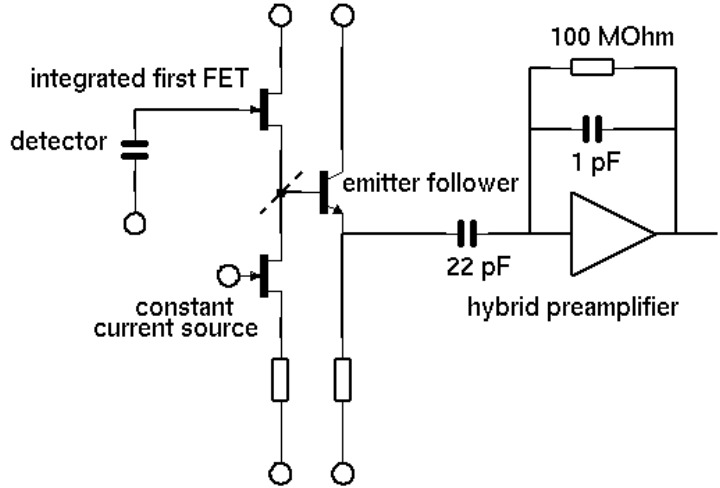}
}
\caption{Our implementation of the SDD readout architecture}
\label{SDDReadout}
\end{figure}

\subsection{Readout Architecture}

\begin{figure}[htbp]
\centerline
{
 \includegraphics[width=2in,angle=-90,clip]{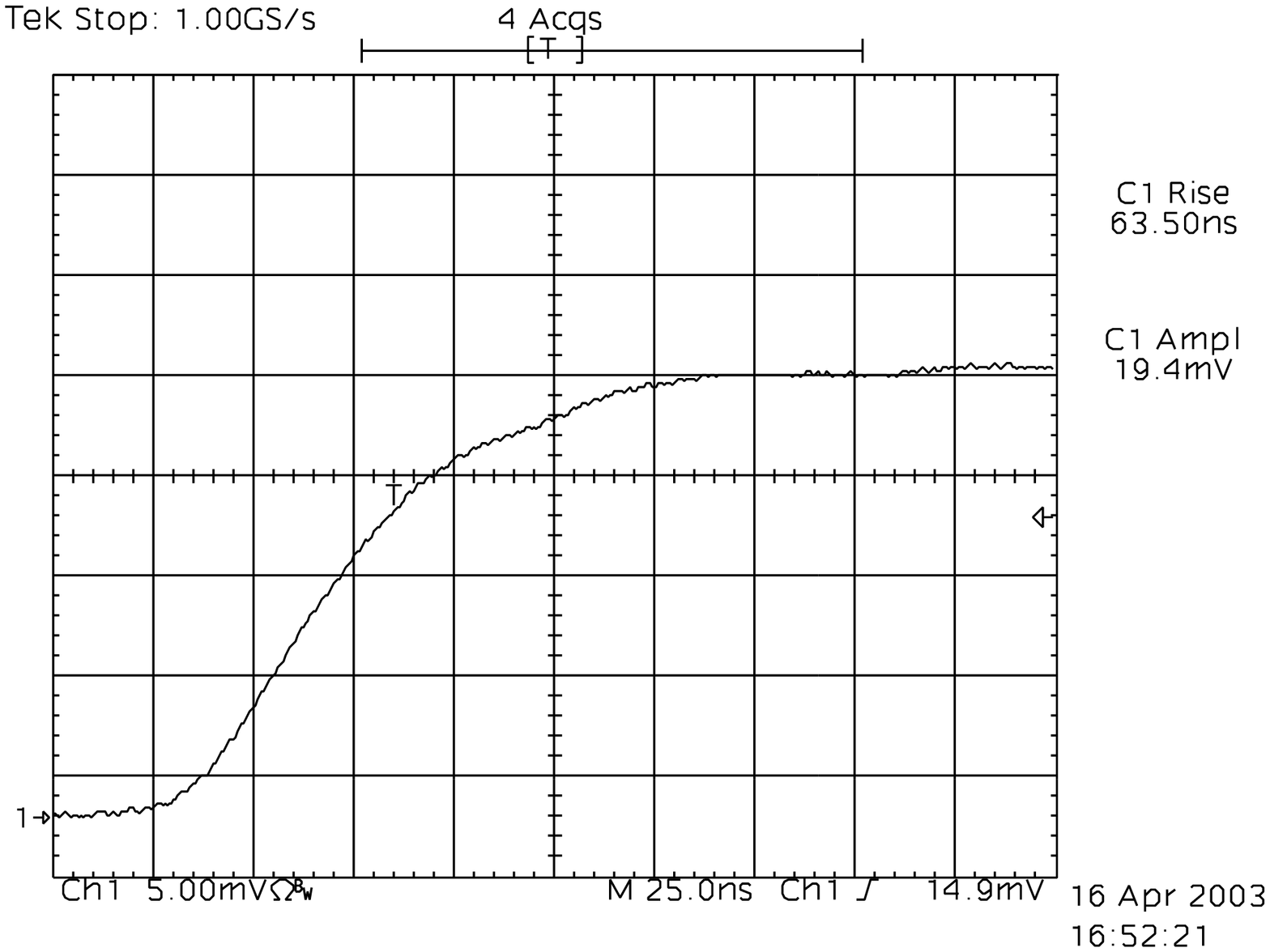}
 \includegraphics[width=2in,angle=-90,clip]{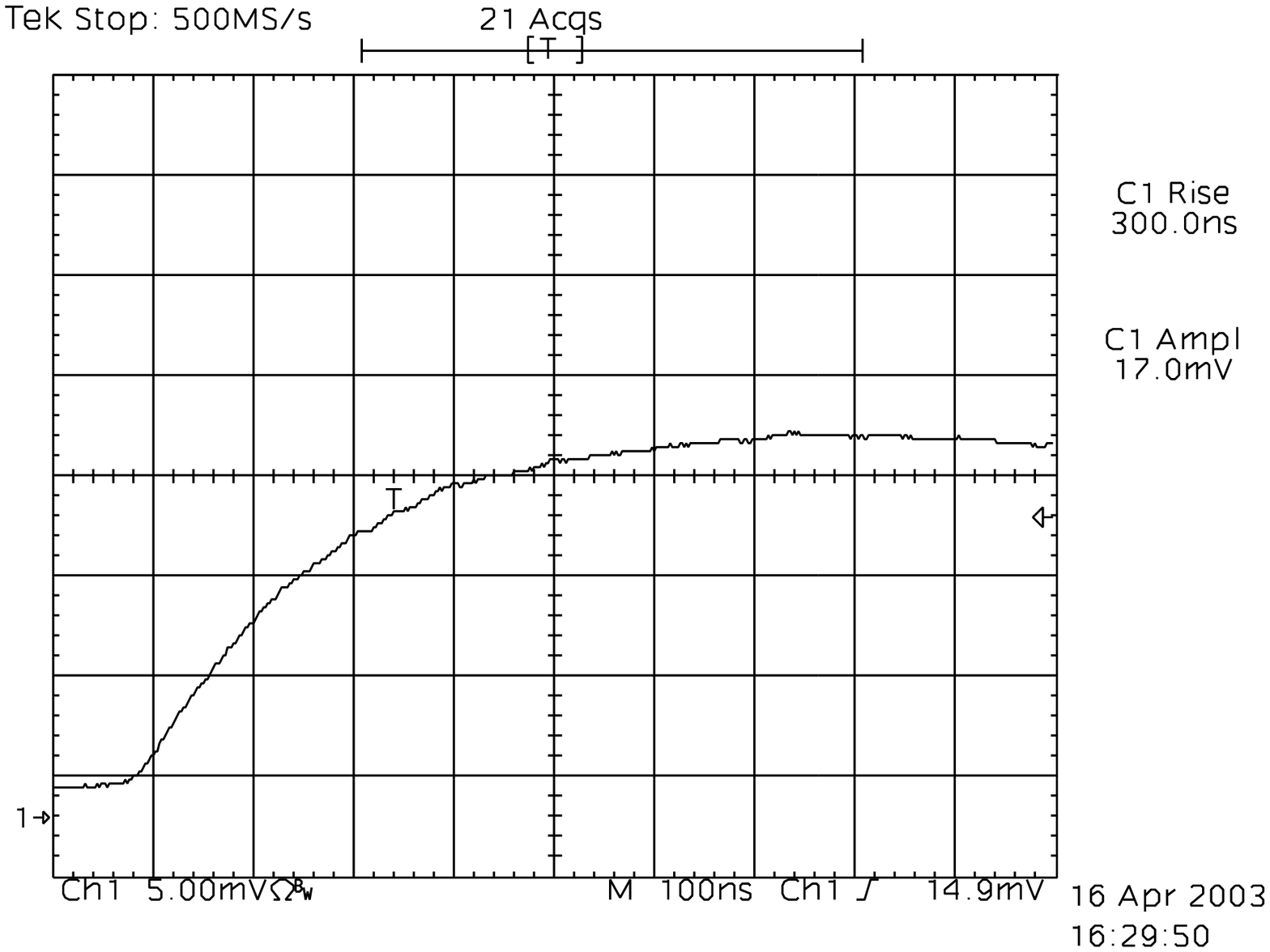}
}
\caption{The rise time of the SDD signal after preamplifier with and without emitter follower stage}
\label{emitterfollower}
\end{figure}

The standard readout architecture for the SDD consists of a source follower with a source load of a constant current supply and the detector signal is amplified at a voltage-sensitive preamplifier, and then further filtered with a shaper. The transconductance (g$_{m}$) of an n channel on-chip JFET is about 0.3 mS and the time constant $\tau$ = $\frac{1}{g_{m}}$ $\cdot$ C$_{total}$ produces a rise time of $2.2 \cdot \tau$ which is of the order of 300 ns. For the Compton camera application it is important to have fast trigger signals from the first detector. This can be done by a readout of the back side of the detector which is under research at MPI. Our implementation of the readout architecture is shown in Fig \ref{SDDReadout}. Emitter followers were used for gas proportional counters as low noise preamplifiers \cite{Farr}. In addition to the source-follower at the SDD chip, an emitter follower provided certain advantages. First of all, the signal after the emitter-follower stage becomes more immune to additional stray capacitances. The readout electronics can be placed further away from the detector, provided that the emitter-follower is as near as possible to the detector. Furthermore, the rise time of the preamplified signal decreases considerably, which makes the use of a short shaping time ($<$100 ns) possible. The driver transistor for the emitter follower was chosen to be a bipolar junction transistor. The selection of this transistor is based on a dynamic range, a noise contribution and an input capacitance. 
Fig \ref{emitterfollower} shows the preamplifier output with and without emitter follower between the detector and the preamplifier. The detector was irradiated by a $^{55}$Fe source. 
The rise time of the signal reduces almost by a factor of 5 with the emitter follower.

\subsection{Preamplifier}

\begin{figure}[htbp]
\centerline
{
 \includegraphics[width=3in,clip]{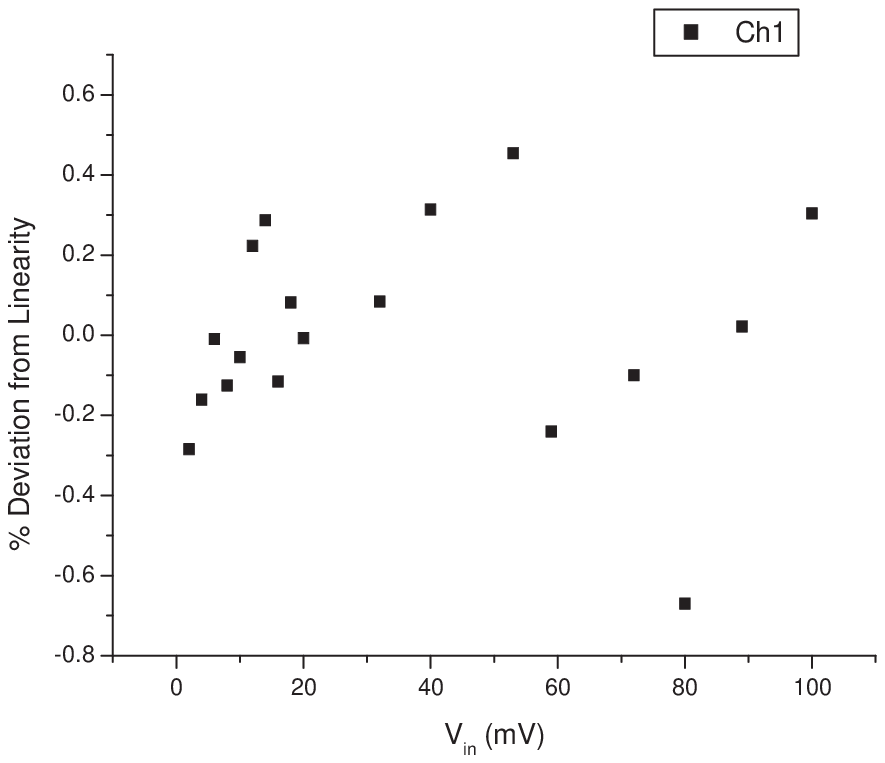}
 \includegraphics[width=3in,clip]{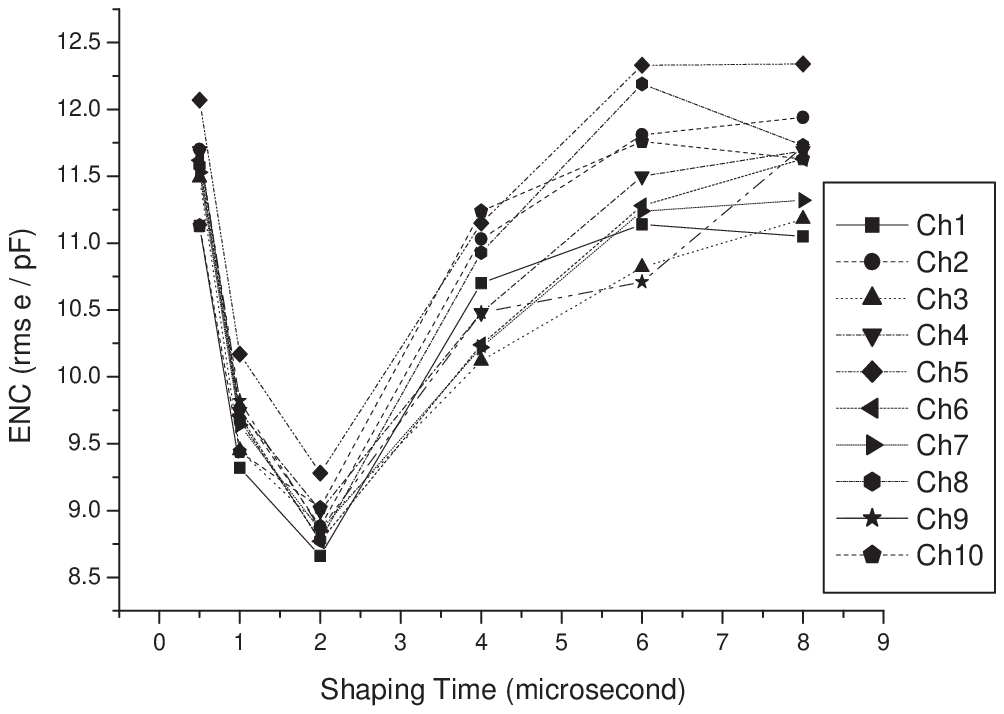}
}
\caption{a) Percent deviation from linearity for the first preamplifier channel b) Noise figure for the preamplifier channels}
\label{PreAmp}
\end{figure}

The preamplifier is a modified version of a preamplifier which was used for one of the first prototype SDDs \cite{CarloESRFpaper}. The  gain of the preamplifier is approximately the ratio between the injection capacitor at the input and the feedback capacitor. The voltage step at the output of the preamplifier decays with a time constant of $R_{f}\cdot C_{f}=100 \mu s$. 
\\
\\
\begin{figure}[htbp]
\centerline
{
 \includegraphics[width=2in,clip]{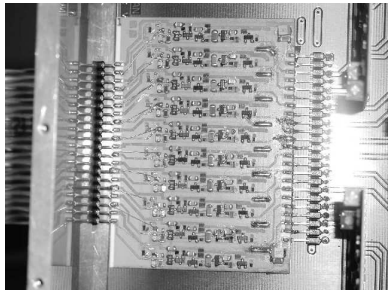}
}
\caption{Hybrid Preamplifier}
\label{Preamplifier}
\end{figure}
10 channels of the preamplifier have been implemented on a hybrid board by using a thick film technology. (fig \ref{Preamplifier}). 
The power consumption per channel is around $130 mW$ according to the SPICE simulations. The deviation from linearity of the preamplifier for one channel is shown in Fig \ref{PreAmp}a. The measurement was done by applying a voltage step at the input of the preamplifier and measuring the output of it. The average voltage gain is around 15. In Fig \ref{PreAmp}b the equivalent noise charge (ENC) of the preamplifier as a function of a shaping time is shown.The preamplifier output was connected to a spectroscopy amplifier (model 1413) for performing the noise measurements. The ENC was calculated for an input capacitance of 1 pF. The peak noise voltage of the preamplifier/shaper system was measured with an analog oscilloscope recording the peak-to-peak amplitude at a very low trigger rate ($\leq$ $100 Hz$) and this level corresponds to four to five standard deviations. The results obtained by this method were also confirmed with a digital oscilloscope's (LeCroy 9362) standard deviation measurement. The ENC values are given in rms electron/pF units which translate into 2-4 rms electrons for 300 fF of detector capacitance. So, the noise contribution of this low noise preamplifier to the whole detector and readout system is expected to be very small.

\subsection{Shaper}

An in house developed hybrid shaper was used for the measurements. The shaper is a CR-RC shaper with adjustable pole-zero cancellation and gain. The hybrid board has two shaper channels and 10 of these boards will be used for the 19-cell SDD setup. 

\begin{figure}[htbp]
\centerline
{
 \includegraphics[width=3.5in,clip]{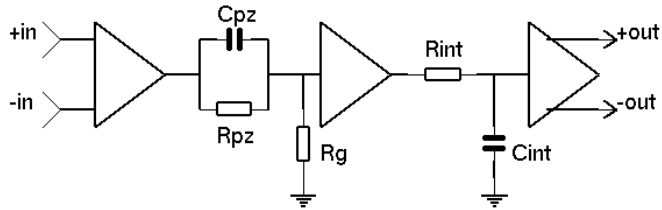}
}
\caption{Block diagram of the shaper}
\label{Shaper}
\end{figure}

\begin{figure}[htbp]
\centerline
{
 \includegraphics[width=3.5in,angle=-90,clip]{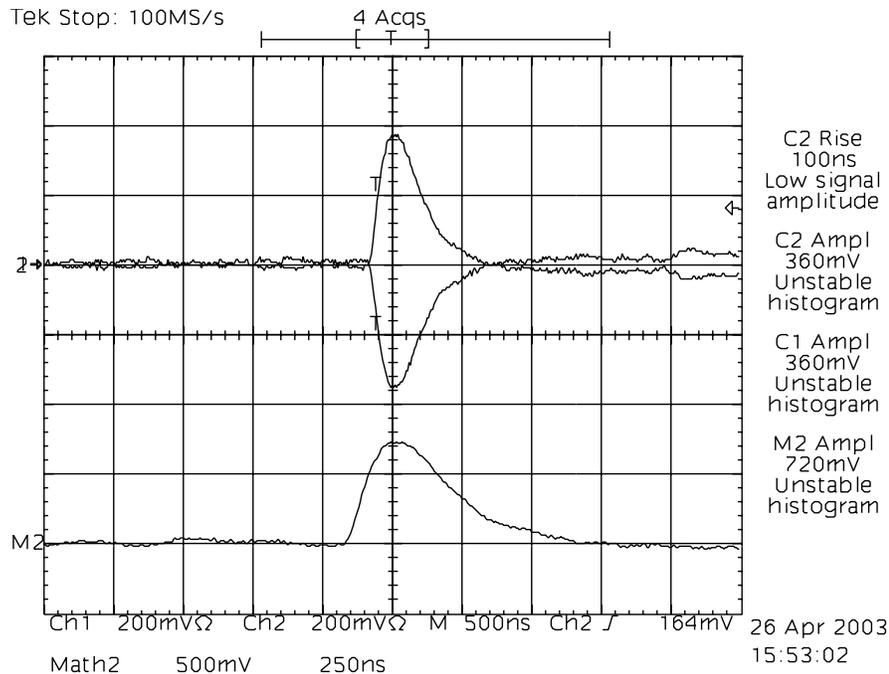}
}
\caption{Shaper output with SDD irradiated by $^{55}Fe$. The upper waveforms are the positive and the negative outputs. Below is the resulting waveform when the difference is taken}
\label{ShaperOutput}
\end{figure}

The block diagram of the shaper is shown in Fig \ref{Shaper}. There is a differential amplifier with variable gain(g) at the first stage. It is followed by a pole-zero(pz) filter and an integrator(int). The waveforms from the shaper, as shown in  Fig \ref{ShaperOutput}, demonstrate reasonably symmetrical differential outputs on a load of 50 $\Omega$. 

Certain aspects were taken into account when choosing the appropriate shaping time. The rise time of the preamplified signal is about 60 ns. The shaping time should not be much shorter than this value in order to avoid ballistic deficit. The sampling rate of the ADC system is 66 MHz and at least 4 samples should be taken at the region of highest signal to noise ratio. The leakage current of this detector is not the dominating factor, therefore the energy resolution improves with an increasing shaping time. However, it can not be made very long, since we need fast signals. Considering these constraints a peaking time of 100 ns was chosen.
 
\subsection{Data Acquisition System}

\begin{figure}[htbp]
\centerline
{
 \includegraphics[width=4.5in,clip]{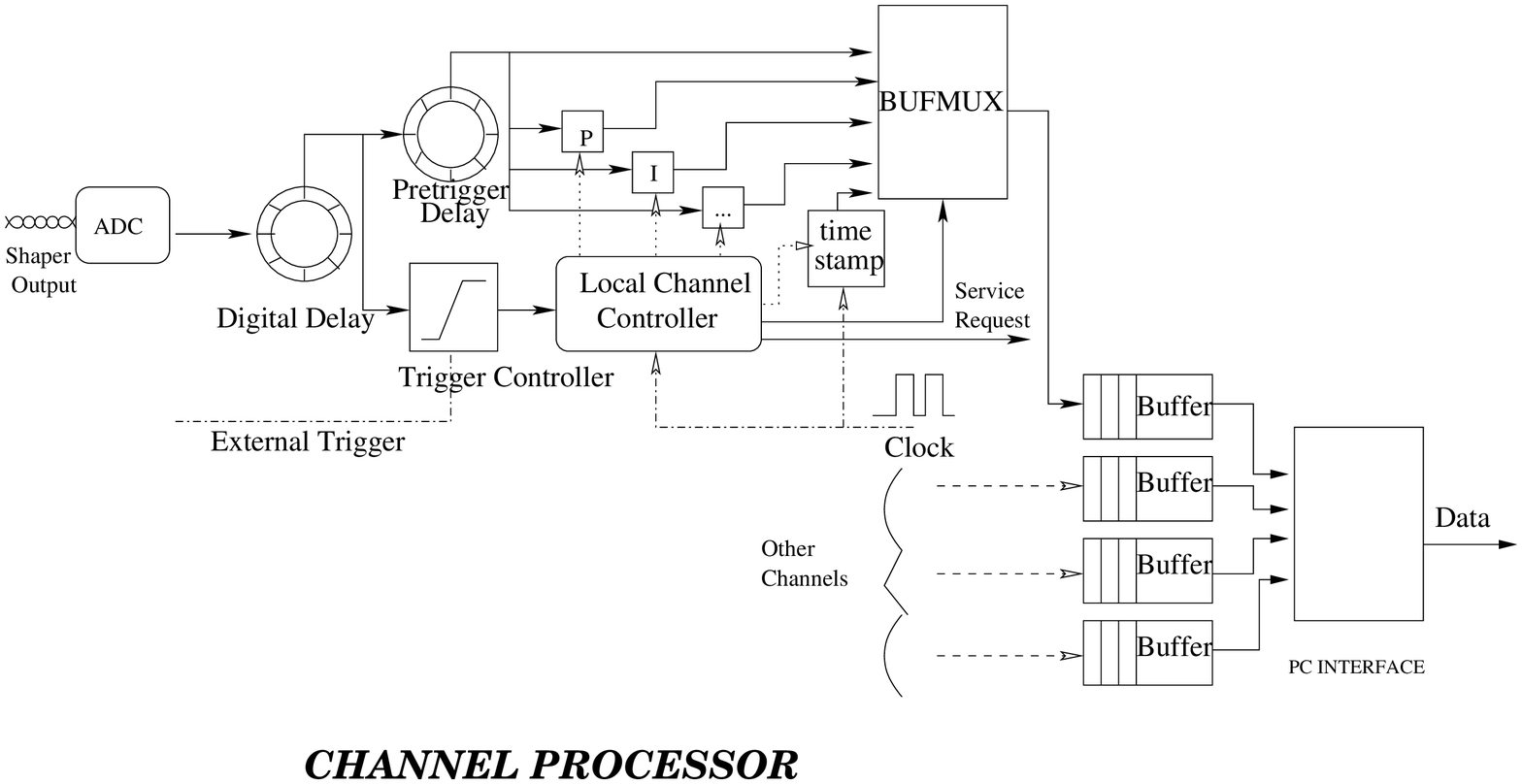}
}
\caption{Data Acquisition for the SDD}
\label{DAQSystem}
\end{figure}

A custom-designed data acquisition (DAQ) system has been developed for the Compton camera prototype \cite{KivancSAMBA}, \cite{KivancISPC}. It consists of channel processors, an event builder and a bus which transfers the data between the channels processors and the event builder. Only one channel of one of the channel processor cards is used for the measurements done with one-cell SDD. The channel processor card is connected to a parallel port interface card and the data are transfered to a computer via this interface. The architecture of the channel processor is shown in Fig \ref{DAQSystem}.

The signals arrive first at the channel processor which has both analog and digital sections. The analog part receives the differential signals, sends them to an ADC and the digitized signals are processed by a Xilinx FPGA. At the FPGA necessary operations such as peak finding, integration and time stamping are applied. 
The ADCs have as mentioned, 66 MHz of sampling rate and 12 Bit resolution. System performance tests show that the effective resolution is around 11 Bit. Each processed signal reaches the interface board and is transfered to a PC. 

\section{Spectroscopic Measurements}

The spectroscopic measurements were done by using the analog and digital readout electronics described at the previous chapters. The differential shaper output is connected to one channel of the single channel processor card. The channel processor is connected to an interface card and the digitized signal is transfered to a PC via parallel port connection. Several bias voltages needed for the SDD are supplied by a custom-designed power supply. 

\begin{figure}[htbp]
\centerline
{
 \includegraphics[width=3in,clip]{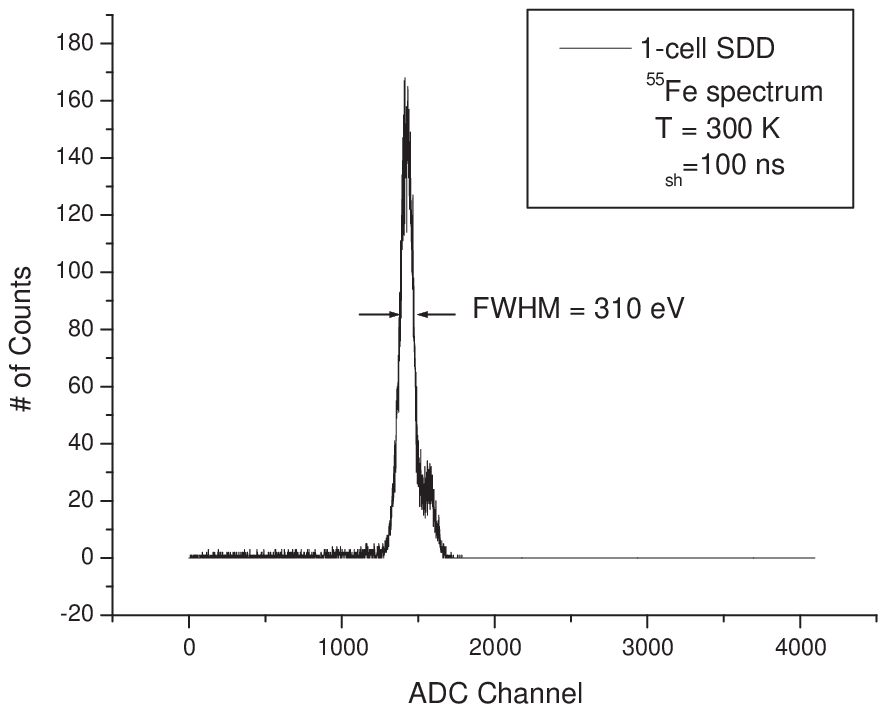}
 \includegraphics[width=3in,clip]{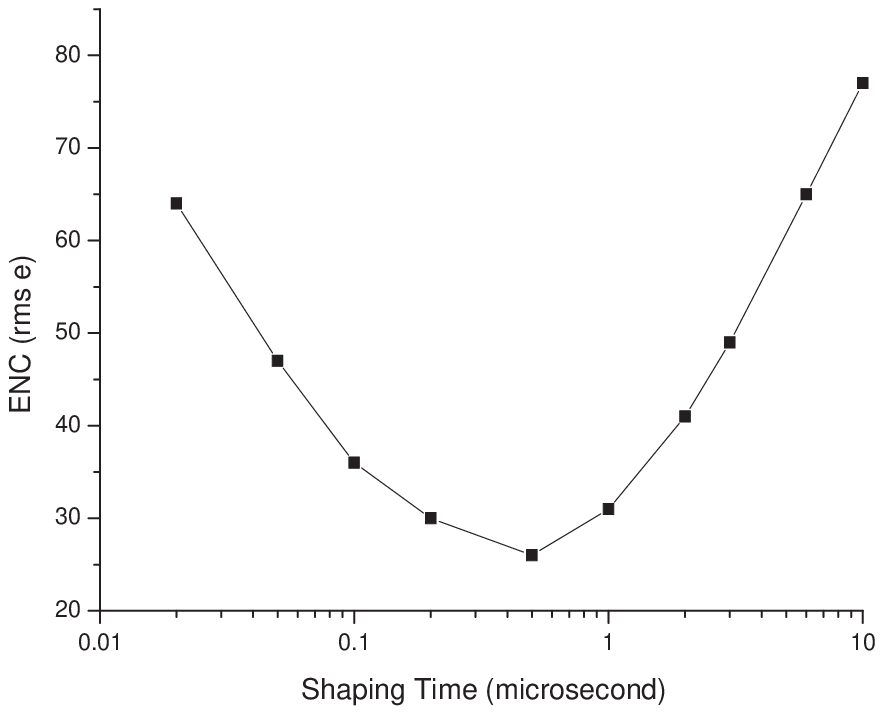}
}
\caption{a) $^{55}$Fe spectrum b) Energy resolution as a function of a shaping time }
\label{SDDFeSpectrum}
\end{figure}

The first measurements were done by using a $^{55}$Fe source. The spectrum is shown in Fig. \ref{SDDFeSpectrum}a, where both Mn K$_{\alpha}$ and K$_{\beta}$ peaks can be seen clearly. The equivalent noise charge (ENC) is about 33 rms electrons which includes noise contributions from the detector and the electronics. Fig \ref{SDDFeSpectrum}b shows the energy resolution at 5.9 keV as a function of a shaping time. The FWHM value is about 250 eV at a shaping time of 250 ns and increases with longer shaping times. 

\begin{figure}[htbp]
\centerline
{
 \includegraphics[width=4.5in,clip]{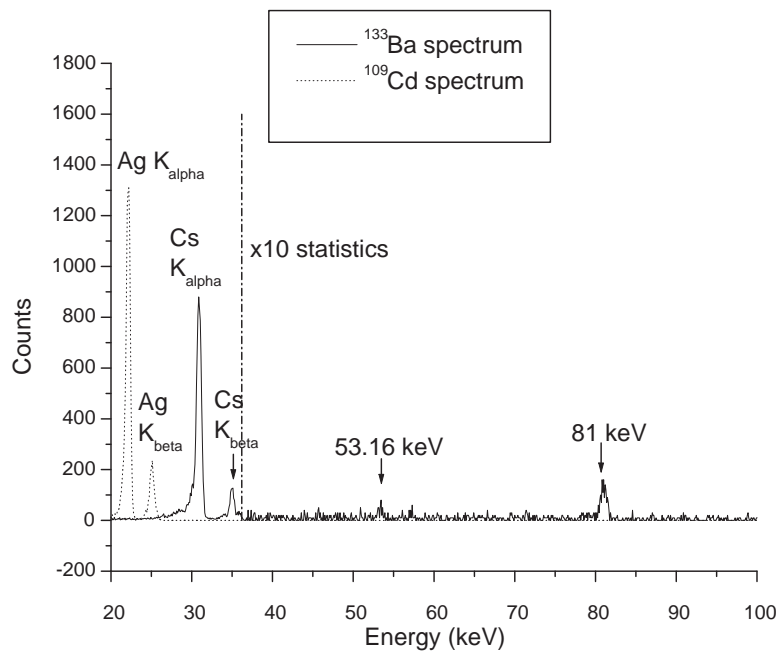}
}
\caption{$^{109}$Cd and $^{133}$Ba spectrums obtained with SDD}
\label{SDDCdBaSpectra}
\end{figure}

 The aim of the Compton camera project is to be able to use high energetic radionuclides, preferably at least few hundered keV and more. Therefore, it is interesting to study the performance of the SDD with a radioactive source emitting higher energetic photons. The spectrums obtained with $^{109}$Cd and $^{133}$Ba sources are shown in Fig. \ref{SDDCdBaSpectra}. The counts are multiplied by 10 for energies above 35 keV in order to see the peaks more clearly. 

\section{Conclusion}

A multi-channel silicon drift detector is going to be used as a scatter detector of a Compton camera prototype. The complete chain of analog and digital readout electronics for the SDD has been tested by using a single cell SDD. Faster readout has been obtained by using an emitter follower stage immediately after the on-chip FET of the detector. A low noise hybrid voltage preamplifier and a CR-RC hybrid shaper have been used. The data acquisition was done by using a custom-designed FPGA based readout system. An ENC of 33 rms electron of the overall system has been measured by using a shaper with a rise time of 100 ns. 

\section{Acknowledgment}

We would like to thank Prof. Lothar Str\"uder from MPI and Siegen for providing us with 1-cell Silicon drift detectors. We are much grateful to Dr.Alan Rudge from CERN and Dr.Helmuth Spieler from BNL for the very useful and instructive discussions. Many thanks go to Karim Laihem for his help in preparing the setup. Last but not the least we thank to Dieter Gebauer, Dieter Junge and Bernd Dostal for the preamplifier production. 





\begin{thebibliography}{00}

\bibitem{Todd} R.W. Todd, J.M. Nightingale and D.B. Everett, ``A Proposed $\gamma$ Camera'', Nature, 251, 132-134, 1974.

\bibitem{Schönfelder} V. Sch\"onfelder, U. Graser and R. Diehl, ``Properties and Performance of the MPI Balloon Borne Compton Telescope'', Astronomy and Astrophysics, 110, 138-151, 1982.

\bibitem{Singh} M. Singh, ``An Electronically Collimated Gamma Camera for Single Photon Emission Computed Tomography. Part 1: Theoretical Considerations and Design Criteria'', Medical Physics, 10, 421-427, 1983.

\bibitem{Martin} J.B. Martin, G.F. Knoll, D.K. Wehe, N. Dogan, V. Jordanov, N. Petrick, M. Singh, ``A Ring Compton Scatter Camera for Imaging Medium Energy Gamma Rays'', IEEE Trans. Nucl. Sci., NS-40, 972-978, 1993.

\bibitem{ourIEEEpaper} T. {\c C}onka Nurdan, K. Nurdan, F. Constantinescu, B. Freisleben, N.A. Pavel, A.H. Walenta, ``Impact of the Detector Parameters on a Compton Camera'', IEEE Trans. Nucl. Sci., Vol:49 Issue:3 Part:1, 817-821, June 2002.

\bibitem{Ott} C.J. Solomon and R.J. Ott, ``Gamma Ray Imaging with Silicon Detectors - A Compton Camera for Radionuclide Imaging in Medicine'', Nucl. Inst. Meth. A 273, 787-792, 1988.

\bibitem{Strüder} L. Str\"uder, P. Lechner and P. Leutenegger, ``Silicon Drift Detector - the key to new experiments'', Naturwissenschaften 85, 539-543, 1998.

\bibitem{MPIHLL} http://www.hll.mpg.de/

\bibitem{Carlos19SDDpaper} C. Fiorini, A. Longoni, F. Perotti, C. Labanti, E. Rossi, P. Lechner, H. Soltau, L. Str\"uder, IEEE Trans. Nucl. Sci., Vol. 49, No. 3, 995, June 2002.

\bibitem{Holography} K. Hansen and L. Tr\"oger, ``A Novel Multicell Silicon Drift Detector Module for X-Ray Spectroscopy and Imaging Applications'', IEEE trans. Nucl. Sci., Vol. 47, No. 6, Dec. 2000.

\bibitem{Spectrometer} C. Fiorini and A. Longoni, ``In-Situ, Non-Destructive Identification of Chemical Elements by Means of Portable EDXRF Spectrometer'', IEEE Trans. Nucl. Sci., Vol. 46, No. 6, Dec 1999.
 
\bibitem{Farr} W.D. Farr and G.C. Smith, ``Emitter Followers as Low Noise Pre-amplifiers for Gas Proportional Detectors'', Nucl. Inst. Meth. 206, 159-167, 1983.

\bibitem{CarloESRFpaper} Ch. Gauthier, J. Goulon, E. Moguiline, A. Rogalev, P. Lechner, L. St\"uder, C. Fiorini, A. Longoni, M. Sampietro, H. Besch, R. Pfitzner, H. Schenk, U. Tafelmeier, A. Walenta, K. Misiakos, S. Kavadias, D. Loukas, ``A High Resolution, 6 Channels, Silicon Drift Detector Array with Integrated JFET's Designed for XAFS Spectroscopy: First X-ray Fluorescence Excitation Spectra Recorded at the ESRF'', Nucl. Inst. Meth. A 382, 524-532, 1996.

\bibitem{KivancSAMBA} K. Nurdan, T. {\c C}onka Nurdan, H.J. Besch, B. Freisleben, N.A. Pavel and A.H. Walenta, ``FPGA Based Data Acquisition System for a Compton Camera'', proceeding of SAMBA (Symposium on Applications of Particle Detectors in Medicine, Biology and Astrophysics) II, Nucl. Inst. Meth. A 510, 122-125, 2003.

\bibitem{KivancISPC} K. Nurdan, H.J. Besch, T. {\c C}onka Nurdan, B. Freisleben, N.A. Pavel and A.H. Walenta, ``Development of a Compton Camera Data Acquisition System Using FPGAs'', proceedings of ISPC (International Signal Processing Conference), 2003. 





\end{thebibliography}
\end{document}